\documentclass[prb,twocolumn,superscriptaddress,showpacs]{revtex4}

\usepackage[T1]{fontenc}
\usepackage[latin9]{inputenc}
\usepackage{amsmath}    
\usepackage{amssymb}
\usepackage{graphicx}   
\usepackage{verbatim}   
\usepackage{color}      
\usepackage{hyperref}   

\makeatletter

\@ifundefined{textcolor}{}
{%
 \definecolor{BLACK}{gray}{0}
 \definecolor{WHITE}{gray}{1}
 \definecolor{RED}{rgb}{1,0,0}
 \definecolor{GREEN}{rgb}{0,1,0}
 \definecolor{BLUE}{rgb}{0,0,1}
 \definecolor{CYAN}{cmyk}{1,0,0,0}
 \definecolor{MAGENTA}{cmyk}{0,1,0,0}
 \definecolor{YELLOW}{cmyk}{0,0,1,0}
}

\makeatother

\usepackage{times}

\begin{document}
\title{Scaling theory vs exact numerical results for spinless resonant level model}

\author{Annam\'{a}ria~Kiss} \email{kiss.annamaria@wigner.mta.hu} 
\affiliation{Institute for Solid State Physics and Optics, Wigner Research Centre for Physics of the Hungarian Academy of Sciences, P. O. B. 49, H-1525, Budapest, Hungary}
\affiliation{BME-MTA Exotic Quantum Phases Research Group, Budapest University of Technology and Economics, Budapest, Hungary}

\author{Junya Otsuki}
\affiliation{Department of Physics, Tohoku University, Sendai, 980-8578, Japan}
\affiliation{Theoretical Physics III, Center for Electronic Correlations and Magnetism, Institute of Physics, University of Augsburg, 86135, Germany}

\author{Yoshio Kuramoto}
\affiliation{Department of Physics, Tohoku University, Sendai, 980-8578, Japan}


\date{\today}
\begin{abstract}

The continuous-time quantum Monte Carlo method is applied to the interacting resonant level model (IRLM) using double expansion with respect to Coulomb interaction $U_{fc}$ and hybridization $V$.  Thermodynamics of the IRLM without spin is equivalent to the anisotropic Kondo model in the low-energy limit. 
Exact dynamics and thermodynamics of the IRLM are derived numerically for a wide range of $U_{fc}$ with a given value of $V$.  
For negative $U_{fc}$, excellent agreement including a quantum critical point is found with a simple scaling formula  that deals with $V$ in the lowest-order, and $U_{fc}$ up to infinite order.
As $U_{fc}$ becomes positive and large, lower order scaling results deviate from exact numerical results.
Possible relevance 
of the results is discussed to certain Samarium compounds with unusual heavy-fermion behavior.

 \end{abstract}
\maketitle

\section{Introduction}

Recently, unusual heavy-fermion state with large specific heat coefficient is found in SmOs$_{4}$Sb$_{12}$, which is almost completely insensitive to external magnetic field\cite{sanada-2005}. 
Furthermore, in several Samarium compounds, intermediate valence is observed experimentally, which is often combined with Kondo-like behavior at low temperatures\cite{smptge}.
These observations suggest that charge degrees of freedom of the $f$ electrons may play a crucial role in formation of the unusual heavy-fermion state in some Sm compounds, and possibly in other systems. 
This situation is best handled by starting from the Anderson model.
In actual systems with valence fluctuations,
the $f$-electron width ($> 10^{2}$K) caused by hybridization with conduction electrons seems much larger 
than the characteristic energy ($\sim$10 K) of the system.
In contrast with Kondo effect that requires the spin degrees of freedom,
we shall search for a charge fluctuation mechanism that gives rise to a smaller energy scale.

Motivated by the situation described above,
we consider the (spinless) interacting resonant level model (IRLM) where an on-site Coulomb interaction term is introduced between the local electron ($f$) and conduction electrons ($c$) at the origin.
The model is given by
\begin{eqnarray}
{\cal H} &=&  {\cal H}_{c} + {\cal H}_{f} + {\cal H}_{\rm hyb} + {\cal H}_{fc} \nonumber\\
&=& 
\sum_{\boldsymbol{k}} \varepsilon_{\boldsymbol{k}} c^{\dag}_{\boldsymbol{k}}c_{\boldsymbol{k}} +  \varepsilon_{f}   f^{\dag}f +
V (f^{\dag}c + c^{\dag}f) \nonumber\\
&+& 
U_{fc} \left( f^{\dag}f - \frac{1}{2} \right) \left( c^{\dag}c - \frac{1}{2} \right),
 \label{eq-ham-0}
\end{eqnarray}
where $c_{\boldsymbol{k}}$ is the annihilation operator of the Bloch state $\boldsymbol{k}$, while 
$c=N^{-1/2}\sum_{\boldsymbol{k}} c_{\boldsymbol{k}}$ with $N$ being the number of sites
denotes the annihilation operator of the Wannier state at the origin.  In this paper we restrict to the case of $\varepsilon_f=0$ and half-filled conduction band.

In the low-energy range, thermodynamics of the IRLM is equivalent to the anisotropic Kondo model as discussed by Vigman-Finkelstein \cite{vigman78} and 
Schlottmann\cite{schlottmann78}.
The IRLM has further been investigated by many authors, and its application to the quantum dot in non-equilibrium has also been made\cite{mehta2006,sela2006}.
In addition, extension to 
multichannels of conduction bands has also been studied by perturbative renormalization group (RG) approach\cite{giamarchi-1993} and numerical renormalization group\cite{borda-2007, borda-2008}.
Recently, a multichannel effect by the assistance of phonons 
was also proposed\cite{ueda-2010}.

In spite of these studies, quantitative information of the model at finite temperature is lacking, especially concerning the dynamics showing crossover to the ground state.   
The dynamics of the IRLM cannot in general be reduced to that of the Kondo model because of different matrix elements of physical quantities.
In this paper we apply the 
continuous-time quantum Monte Carlo (CT-QMC)  
method\cite{gull-2013} to investigate the single-channel IRLM 
for a wide range of the Coulomb interaction.
We pay particular attention to the case of negative $U_{fc}$, which 
includes a quantum critical behavior.
Because of apparently unphysical sign, some interesting aspects with $U_{fc}<0$ has been overlooked. 
Based on the numerical data we can test the applicability of perturbative analytic approaches and the phase shift scheme.
Especially, we are interested in the behavior near the
quantum critical point emerging in the negative $U_{fc}$ range, where the renormalized hybridization vanishes. 

This paper is organized as follows. 
In Section~\ref{section-weakc} we rederive
the renormalized hybridization by the perturbative RG approach for weak-coupling regime of both $V$ and $U_{fc}$, and then summarize the phase shift scheme 
for larger $U_{fc}$, keeping $V$ small.
In Section~\ref{section-ctqmc} the CT-QMC algorithm is formulated and its details are discussed.
Numerical results for the IRLM are given in Section~\ref{section-numresults},
emphasizing the dynamical property at finite temperatures.
Finally, Section~\ref{section-discussion} is devoted to discussion and the summary of this paper.

\section{Analytic results for perturbative renormalization }\label{section-weakc}

\subsection{Renormalization of hybridization}

There are many analytical methods to take account of simultaneous effects of hybridization $V$ and Coulomb interaction $U_{fc}$, such as Bethe ansatz\cite{filyov}, bosonization \cite{bosonization}, mapping to Anderson-Yuval Coulomb gas\cite{borda-2007, borda-2008, Yuval-1970}, and scaling\cite{schlottmann-I, schlottmann-II}.
We find it most compact to 
use the effective Hamiltonian method\cite{kuramoto-2000}. In this approach a model space is introduced which contains only a part of the original Hilbert space.
If $\psi_{i}$ and $E_{i}$ are eigenstates and eigenvalues of the original problem as ${\cal H} \psi_{i} = E_{i} \psi_{i}$, than we require that the same eigenvalues, although only a part of the original ones, are reproduced within the model space by the effective Hamiltonian as ${\cal H}_{\rm eff} P\psi_{i} = E_{i} P\psi_{i}$, where $P$ is the projection operator to the model space.
The effective Hamiltonian is constructed in lowest orders within the Rayleigh-Schr\"odinger perturbation theory as\cite{kuramoto-2000}
\begin{eqnarray}
{\cal H}_{\rm eff} &=& P({\cal H}_{0}+{\cal H}_1)P + {\cal H}_1 \frac{1}{\varepsilon_{i}-{\cal H}_{0}} Q {\cal H}_1 \nonumber\\
&+&  {\cal H}_1 \frac{1}{\varepsilon_{i}-{\cal H}_{0}} Q  {\cal H}_1 \frac{1}{\varepsilon_{i}-{\cal H}_{0}} Q {\cal H}_1\nonumber\\
&-& \sum_{j}  {\cal H}_1 \frac{1}{\varepsilon_{i}-{\cal H}_{0}}  \frac{1}{\varepsilon_{j}-{\cal H}_{0}}  Q {\cal H}_1 |j \rangle \langle j | {\cal H}_1,\label{eq-effhyb}
\end{eqnarray}
where $Q=1-P$, $\varepsilon_{i}$ is the energy of the initial conduction electron state, and the original Hamiltonian is divided as ${\cal H}={\cal H}_{0}+{\cal H}_1$
with ${\cal H}_1$ including both $U_{fc}$ and $V$ terms.

The renormalization procedure is performed by reducing the model space starting from the original Hilbert space.
Namely, the conduction electron states near the band edges are disregarded, i.e. $Q$ is chosen as a projection to a space with conduction electron states within the ranges $[-D,-D+\delta D]$ and $[D-\delta D,D]$, where $\delta D$ is infinitesimal. During this procedure the bare interactions are modified and will depend on the new cut-off energy  $D^{\prime}=D-\delta D$.
\begin{figure}
\centering
\includegraphics[width=0.8\hsize]{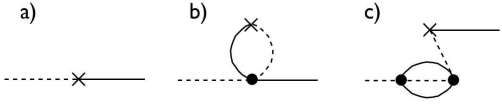}
\caption{Diagrams in leading and next-leading orders contributing to the renormalization of the hybridization. {\sl Solid line} represents conduction electron state, while {\sl dashed line} corresponds to the (local) impurity state. {\sl Cross} indicates the hybridization process and the {\sl filled dot} the Coulomb interaction.}
\label{fig-V-diagrams}
\end{figure}
The diagrams shown in Fig.~\ref{fig-V-diagrams} should be considered in leading and next-leading orders of ${\cal H}_1$.
The panels a), b), and c) correspond to the first, second and fourth term in Eq.~(\ref{eq-effhyb}), respectively.
The diagram c) is an example of the "folded digram" which enables automatic consideration\cite{kuramoto-2000} of a model state $j$.

Evaluating the diagrams shown in Fig.~\ref{fig-V-diagrams} based on Eq.~(\ref{eq-effhyb}) we obtain the renormalized hybridization as
\begin{eqnarray}
V^{\prime} = V \left( 1 - u \frac{\delta D}{D} + \frac{1}{2} u^2 \frac{\delta D}{D}  \right),\label{eq-effhyb1}
\end{eqnarray}
where we introduced $u=\rho_{0}U_{fc}$
with $\rho_0$ being the density of conduction band states.
Writing $\delta V = V^{\prime}-V$ in Eq.~(\ref{eq-effhyb1}) and  integrating both sides we obtain
\begin{eqnarray}
V^{\prime}(D^{\prime}) = V \left[\frac{D^{\prime}}{D} \right]^{-u + u^2/2}.\label{eq-effhyb2}
\end{eqnarray}
Equation (\ref{eq-effhyb2})
gives a relation between $V'$ and $D'$, but does not give the renormalized hybridization in terms of bare parameters $V, U_{fc}$ and $D$.
We follow Borda {\it et al.}\cite{borda-2007} to impose a self-consistent condition. Namely, we stop
the renormalization process of $D'$ at 
the resonance width 
 $\Delta^{\prime}
 =\pi \rho_{0}(V^{\prime})^2$. 
Then the renormalized hybridization $V^*$ is given in terms of 
the bare parameters as
\begin{eqnarray}
V^{*} = V  \left[ \frac \Delta D \right]^{(-u+u^2/2)/(1+2u-u^2)}
\label{eq-effhyb3}
\end{eqnarray}
with $\Delta \equiv \pi \rho_{0} V^2$.
For later purpose, we also quote another form
that is equivalent to Eq.(\ref{eq-effhyb3}):
\begin{align}
\ln \frac {\Delta^*} \Delta = -\eta \ln \frac{\Delta^*} D,
\label{equivalent}
\end{align}
where 
$\Delta^* \equiv \pi \rho_{0} (V^*)^2$ and
\begin{align}
\eta \equiv u(2-u).
\label{eta}
\end{align}

We remark  
that the  Coulomb interaction is not renormalized up to\cite{nozieres-1969-I, nozieres-1969-II} ${\cal O}({\cal H}_1^3)$, i.e. $U_{fc}^{\prime} = U_{fc}$.

\subsection{Vanishing hybridization at critical $U_{fc}$}

The exponent $x(u)$ in Eq.~(\ref{eq-effhyb3}) is written as
\begin{align}
x(u) \equiv \frac{-u+u^2/2}{1+2u-u^2} = -\frac 12 -
\frac 1{2(u-u_-)(u-u_+)}
\end{align}
with
\begin{eqnarray}
u_{\pm}=1\pm \sqrt{2}.
\label{eq-ucritical}
\end{eqnarray}
Namely, we obtain $x(u)\sim -u$ for $|u|\ll 1$, and divergent $x(u)$ to positive infinity as $u$ approaches $u_{\pm}$ in the range $u_-<u<u_+$.
If we take the result literally, we expect
$V^{\ast}\rightarrow 0$ as $u\rightarrow u_{\pm}$ with $\Delta/D<1$.
Since the perturbative renormalization can be justified only for small $|u|$, we have to be cautious about the result with $|u|\sim {\cal O}(1)$.

It is interesting to compare with the mapping of the IRLM to the anisotropic Kondo model\cite{schlottmann-kondo-I}.  We obtain 
the correspondence:
\begin{align}
&\rho_{0}J_{\perp} = 2V, \\
&\rho_{0}J_{\|} = 
\sqrt{2}\left( \rho_{0}U_{fc} + \sqrt{2} - 1 \right) = 
\sqrt{2}\left( u - u_- \right)  .
\end{align}
Namely, the critical value $u=u_-$ corresponds to 
$J_{\|}=0$ 
in the Kondo model.
If $|J_{\perp}|$ is negligibly small, the point 
$J_{\|}=0$ separates the singlet and doublet ground states in the Kondo model, and gives a quantum critical point.  In the IRLM, the quantum critical point corresponds to degeneracy of vacant and occupied states at 
$V'=0$ with $\varepsilon_f=0$.
One may naturally ask why the weak-coupling renormalization and bosonization gives precisely the same result in the strong-coupling region.
Since both the perturbative RG and bosonization are weak-coupling theories, coincidence of  the results does not guarantee the correct behavior around $u \sim u_{-}={\cal O}(1)$.
In the following, we derive the correct value in terms of a phase shift, which is indeed very different from $u_-\sim -0.41$.

Let us assume negligible $V$ in the IRLM, but large value of $|U_{fc}|$.
Then we 
can utilize
the analogy with the x-ray threshold problem\cite{nozieres-1969-III}. Namely, 
neglecting the interference between $U_{fc}$ and $V$, 
the Coulomb interaction is replaced by the phase shift as
\begin{eqnarray}
u = \rho_{0}U_{fc} \,\, \longrightarrow \,\, 
\tilde{u} = \frac{\delta_{U}}{\pi} = \frac{1}{\pi} {\rm arctan} (\pi \rho_{0} U_{fc}),
\label{eq-phsufc}
\end{eqnarray}
which takes account of multiple scattering by $U_{fc}$ to infinite order without, however, considering intervening hybridization.
This phase shift scheme should work well for $0> \tilde{u} \gtrsim u_-$ since the renormalized hybridization becomes negligible in this case.
Using the condition $1+2\tilde{u}-\tilde{u}^2=0$ together with Eq.~(\ref{eq-phsufc}), we obtain
\begin{eqnarray}
u_{\rm cr} \equiv \rho_0 U_{fc}^{\rm cr} = -\frac{1}{\pi}\tan(\sqrt{2}\pi) \approx -2.3,
\label{eq-ucritical2}
\end{eqnarray}
which is the (single) critical value $U_{fc}^{\rm cr}$ in the phase shift scheme. 
We expect the phase shift description to be exact in the limit of small $V$.
With a finite bare hybridization, however, the value given by Eq.~(\ref{eq-ucritical2}) will not be exact.
We shall show later that numerical results nevertheless are in fair agreement with Eq.~(\ref{eq-ucritical2}) even for $V\sim 0.3$. 
Then we are led to the formula of renormalized hybridization in the phase shift scheme:
\begin{eqnarray}
V^{*} = V  \left[ \frac \Delta D \right]^{(-\tilde{u}+\tilde{u}^2/2)/(1+2\tilde{u}-\tilde{u}^2)},
\label{V^*}
\end{eqnarray}
which is obtained from Eq.~(\ref{eq-effhyb3}), and improves it for negative $U_{fc}$.

For positive $U_{fc}$, on the other hand,
the phase shift description leads to saturation
$
\tilde{u} \rightarrow 1 <u_+
$
for large $\tilde{u}$.  
Hence 
$V^{\ast}$ remains finite for any $\tilde{u} >0$ instead of vanishing at $\tilde{u}=u_+$.

\subsection{Characteristic energy scale at finite temperature}

By analogy with the Kondo problem, we can define a characteristic energy scale $\Delta^*$ given by Eq.~(\ref{equivalent}), which corresponds to
the halfwidth at half-maximum of the renormalized resonance peak.
This scale also defines a characteristic temperature $T^{\ast}=\Delta^*$
where we set $k_{\rm B}=1$.
In order to compare 
analytic results with numerical results at finite temperature $T$, we follow 
the argument of Schlottmann\cite{schlottmann-II} and use the replacement in Eq.~(\ref{equivalent}):
\begin{align}
-\ln \frac{\Delta^*} D \rightarrow \ln\frac{D}{2\pi T} -\psi
\left( \frac{1}{2} + \frac{\Delta^*}{2\pi T}  \right),
\end{align}
where we use the digamma function $\psi (z) \equiv \Gamma '(z)/\Gamma(z)$.
It can be checked that the limit of $T\rightarrow 0$ recovers Eq.~(\ref{equivalent}).
Then we obtain
\begin{align}
\ln\frac{\Delta^*}{\Delta} = \eta \left[ 
\ln\frac{D}{2\pi T} -\psi
\left( \frac{1}{2} + \frac{\Delta^*}{2\pi T}  \right)
 \right],
\label{eq-deltapt}
\end{align}
which determines $\Delta^*=\Delta^*(T)$ at finite temperature.

Furthermore, we define a characteristic value $U_{fc}^{\ast}(T)$ of the Coulomb interaction
at a given temperature $T$ by
the condition 
\begin{align}
\Delta^{*} (T=0, U_{fc}^{\ast}) 
= T.
\end{align}
Then, use of Eq.~(\ref{equivalent}) gives the corresponding $\eta =\eta^*$ by
\begin{eqnarray} 
\eta^{\ast} (T)= \frac{{\rm ln} (T/\Delta) }
{{\rm ln} (D/T) }.
\end{eqnarray}
We obtain now $u^*\equiv \rho_0U_{fc}^{\ast}$ explicitly 
as the solution of Eq.~(\ref{eta}) 
with $\eta = \eta^{\ast} $.
In the phase shift scheme,
the result is given by
\begin{eqnarray}
\pi u^* = \pi \rho_{0} U_{fc}^{\ast}(T) = 
 {\rm tan} 
\left[  \pi \left( 1 - \sqrt{1 - \eta^* } \right)\right] ,
\label{eq-charufc}
\end{eqnarray}
which gives the corresponding phase shift:
\begin{eqnarray}
\delta_{U}^{\ast} \equiv {\rm arctan} (\pi \rho_{0} U_{fc}^{\ast}).
\label{delta_U}
\end{eqnarray}

\section{Monte Carlo Method}\label{section-ctqmc}

\subsection{Treatment of the Coulomb interaction}

In this section, we present an algorithm of the CT-QMC\cite{gull-2013} to treat the model~(\ref{eq-ham-0}).
We begin with the hybridization-expansion algorithm (CT-HYB)\cite{werner-PRL-2006, werner-PRB-2006} and consider how to include $U_{fc}$.  
Although a direct expansion with respect to $U_{fc}$ (CT-INT)\cite{rubtsov-2005} is rather straightforward for this model,
the algorithm based on the CT-HYB brings an advantage in extending the method to the multichannel case. Since the matrix element for the $f$ state is easily taken into account in the CT-HYB, increasing the channel number does not produce additonal cost concerning the evaluation of the $f$ part.

Before proceeding to detailed descriptions, we rewrite the Hamiltonian in Eq.~(\ref{eq-ham-0}) as
${\cal H} =  \widetilde{\cal H}_{c} + \widetilde{\cal H}_{f} + {\cal H}_{\rm hyb} + \widetilde{\cal H}_{fc}$ with the interaction term
\begin{align}
\widetilde{\cal H}_{fc} 
= U_{fc} \left( f^{\dag}f - \alpha_{f} \right) \left( c^{\dag}c - \alpha_{c} \right).\label{eq:tildehfc}
\end{align}
The parameters $\alpha_f$ and $\alpha_c$ are introduced to avoid negative weight configurations,\cite{rubtsov-2005} which will be discussed later.
Correspondingly, ${\cal H}_c$ and ${\cal H}_f$ are rewritten as 
$\widetilde{\cal H}_{c}={\cal H}_{c} + (\alpha_f-1/2) U_{fc} c^{\dag}c$ and 
$\widetilde{\cal H}_{f}={\cal H}_{f} + (\alpha_c-1/2) U_{fc} f^{\dag}f$, respectively.

\begin{figure}
\centering
\includegraphics[width=0.75\hsize]{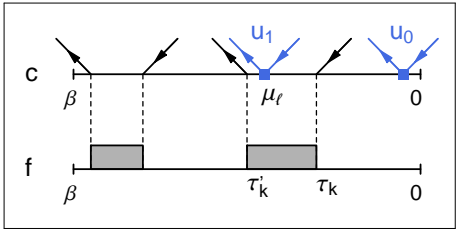}
\caption{
(Color online) An example of the Monte Carlo configuration of order $q=m=2$.
Incoming (outgoing) arrow shows the annihilation (creation) of a conduction electron.
The Coulomb interaction process is indicated by {\sl box symbols}. 
}
\label{fig-ctqmc}
\end{figure}

We begin with the partition function ${\cal Z}$ in a form for expansion with respect to ${\cal H}_{\rm hyb}$ and $\widetilde{\cal H}_{fc}$
\begin{eqnarray}
{\cal Z} &=& \text{Tr} T_{\tau} e^{-\beta (\widetilde{\cal H}_c + \widetilde{\cal H}_f)} \nonumber\\
&\times&
\exp \left\{ -\int_0^{\beta} d\tau [ {\cal H}_{\rm hyb}(\tau) + \widetilde{\cal H}_{fc}(\tau) ]\right\}.
\label{partf1}
\end{eqnarray}
Since the occupation number $n_f$ of the $f$ state is conserved by $\widetilde{\cal H}_c + \widetilde{\cal H}_f$, the ``segment picture"\cite{werner-PRL-2006} can be used for evaluation of the trace for $f$ operators. 
The $f$ state $|f(\tau) \rangle$ fluctuates between the empty state $|0 \rangle$ and the occupied state $|1 \rangle$ by the hybridization term ${\cal H}_{\rm hyb}$ as shown in Fig.~\ref{fig-ctqmc}.
On the other hand, the interaction $\widetilde{\cal H}_{fc}$ does not change $|f(\tau) \rangle$.
Hence, for a given configuration of segments, i.e., for a fixed $|f(\tau) \rangle$, $\widetilde{\cal H}_{fc}(\tau)$ can be regarded as a scattering
\begin{align}
\widetilde{\cal H}_{fc}(\tau) 
= u(\tau) \left[ c^{\dag}(\tau) c(\tau) -\alpha_c \right],
\end{align}
by a ``time-dependent potential"
\begin{align}
u(\tau) 
&= \begin{cases}
-\alpha_f U_{fc} \equiv u_{0} & \text{for $|f(\tau)\rangle=|0 \rangle$} \\
(1-\alpha_f) U_{fc} \equiv u_{1} & \text{for $|f(\tau)\rangle=|1 \rangle$}
\label{time-dep-pot}
\end{cases}.
\end{align} 
Although it is, in principle, possible to integrate out $\widetilde{\cal H}_{fc}(\tau)$ for a fixed $|f(\tau) \rangle$, it is expensive to do it every time $|f(\tau) \rangle$ is changed during simulations.
Instead, we expand with respect to $\widetilde{\cal H}_{fc}$ as well as ${\cal H}_{\rm hyb}$.

By performing the double expansion, the partition function in Eq.~(\ref{partf1}) is expressed as
\begin{eqnarray}
\frac{\cal Z}{{\cal Z}_0} &=& \sum_{q=0}^{\infty} \sum_{m=0}^{\infty} \int d\boldsymbol{\tau}  \int d \boldsymbol{\mu} 
V^{2q} (-1)^{m} u(\mu_{1}) \dots u(\mu_{m}) \nonumber\\
&\times&
W_{c}(\boldsymbol{\tau}, \boldsymbol{\mu}) W_{f}(\boldsymbol{\tau}),\label{partf2}
\end{eqnarray}
where ${\cal Z}_0$ denotes the partition function for $V=U_{fc}=0$.
$\boldsymbol{\tau} = \{ \tau_{1}, \tau_{1}^{\prime}, \dots, \tau_{q}, \tau_{q}^{\prime} \}$ and $\boldsymbol{\mu}=\{ \mu_{1}, \dots, \mu_{m} \}$ are the imaginary times where the hybridization and Coulomb scattering take place, respectively.
Figure~\ref{fig-ctqmc} shows an example of the configuration of order $q=m=2$.
The weight $W_{f}(\boldsymbol{\tau})$ is the thermal average of $f$ and $f^{\dag}$ operators with respect to ${\cal H}_f$, and is the same as in the case of the Anderson model\cite{werner-PRL-2006, werner-PRB-2006}.
The weight $W_{c}(\boldsymbol{\tau}, \boldsymbol{\mu})$ incorporates $c$ and $c^{\dag}$ operators which arise from ${\cal H}_{fc}$ and ${\cal H}_{\rm hyb}$.
The Wick theorem reduces the thermal average to the determinant, 
$W_{c}(\boldsymbol{\tau}, \boldsymbol{\mu})=\det D(\boldsymbol{\tau}, \boldsymbol{\mu})$,
with $D$ being a $(q+m) \times (q+m)$ matrix composed of four blocks:
\begin{align}
D(\boldsymbol{\tau}, \boldsymbol{\mu}) 
= \left( \begin{array}{c|c}
\tilde{g}(\tau_i - \tau_j') & \tilde{g}(\tau_i - \mu_j) \\
\hline
\tilde{g}(\mu_i - \tau_j') & \tilde{g}(\mu_i - \mu_j - 0) - \alpha_c \delta_{ij}
\end{array} \right),
\end{align}
where $\tilde{g}(\tau)$ is the Fourier transform of 
$\tilde{g}(i\omega_n)=[g(i\omega_n)^{-1} - (\alpha_f-1/2) U_{fc}]^{-1}$ with 
$g(i\omega_n)=N^{-1} \sum_{\boldsymbol{k}}(i\omega_n-\varepsilon_{\boldsymbol{k}})^{-1}$.

\begin{figure}
\centering
\includegraphics[width=0.9\hsize]{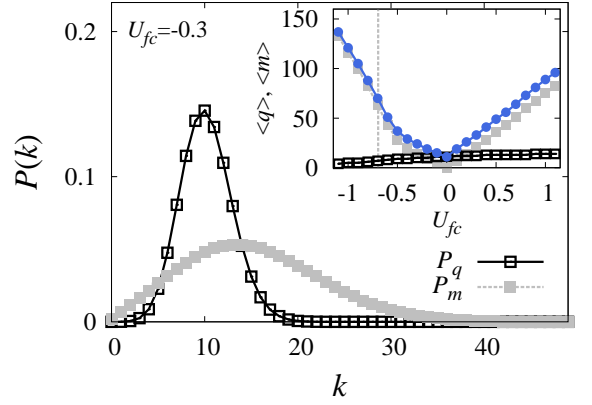}
\caption{(Color online) The probability distribution $P_q$ ($P_m$) of the $V$-expansion ($U_{fc}$-expansion) order with $U_{fc}=-0.3$, $V=0.316$, $T=0.01$. The {\sl inset} shows the average expansion orders, $\langle q \rangle$ and $\langle m \rangle$, as a function of $U_{fc}$. The average matrix size $\langle q+m \rangle$ is also shown as {\sl blue dots}. The vertical {\sl dashed line} indicates the characteristic Coulomb interaction $U_{fc}^{\ast}$. 
Constant density of states is used for the conduction electrons, and the technical parameters are chosen as  $\alpha_c=1.5$, $\alpha_f=1-\delta/U_{fc}$ with $\delta=0.01$.}
\label{fig-prob}
\end{figure}

\begin{figure*}
\centering
\includegraphics[width=0.33\hsize]{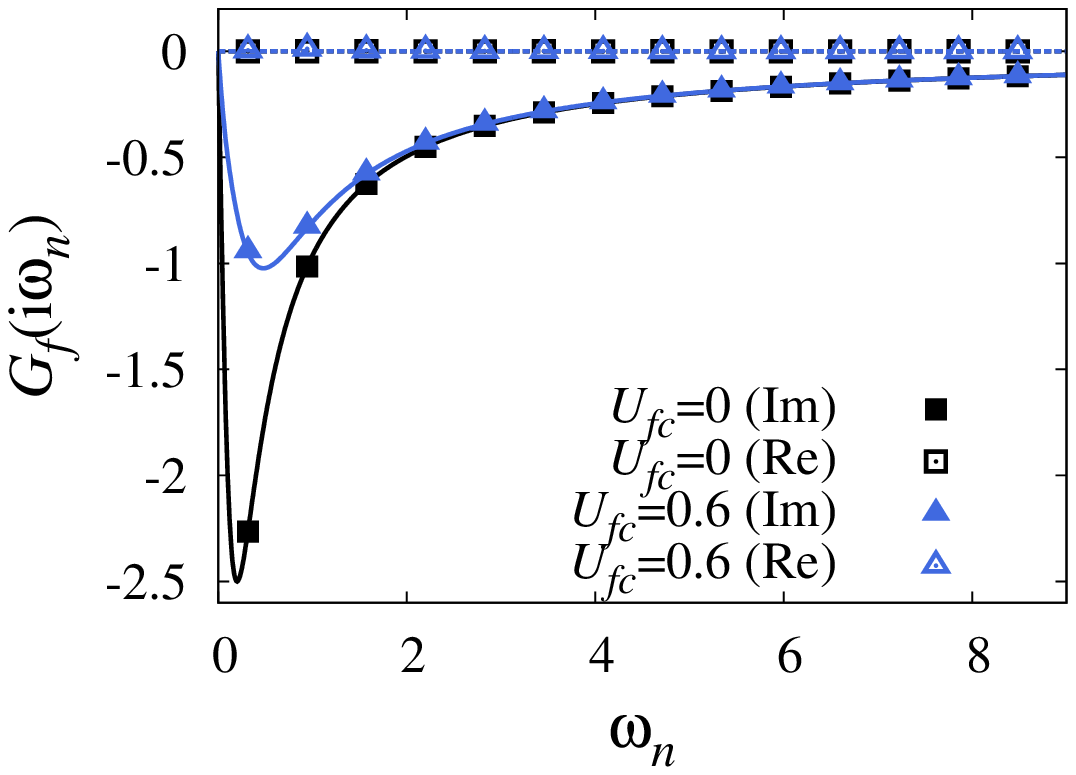}
\includegraphics[width=0.33\hsize]{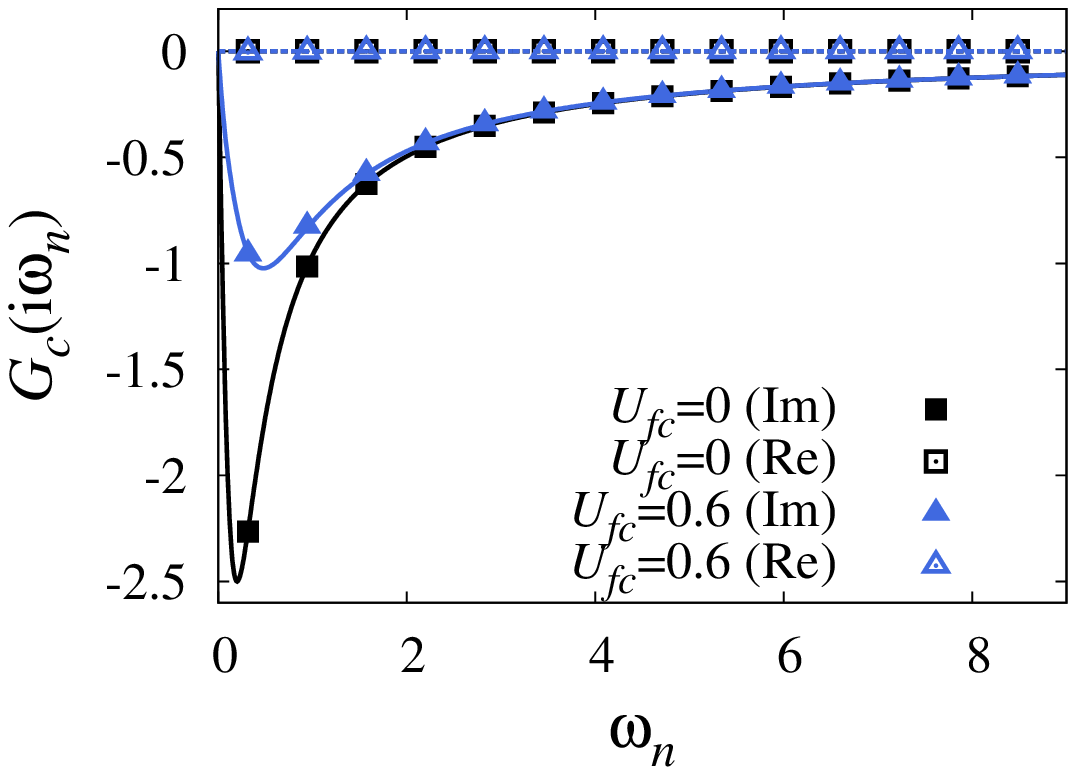}
\includegraphics[width=0.33\hsize]{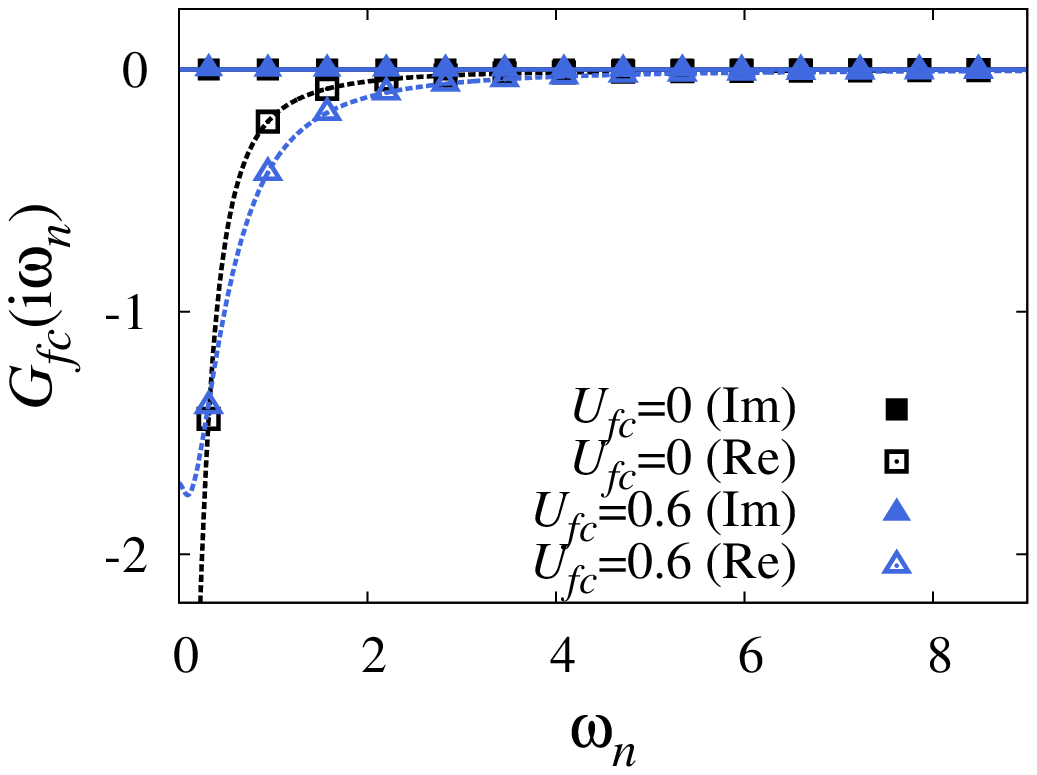}
\caption{(Color online) Comparison of the Green's functions $G_f$, $G_c$ and $G_{fc}$ for the toy model computed in the CT-QMC ({\sl symbols}) with the exact results ({\sl lines}, {\sl dashed:} Re, {\sl solid:} Im). 
The parameter values are chosen as $V=0.2$, $T=0.1$, and $\varepsilon_c=\varepsilon_f=0$ which corresponds to the particle-hole symmetric case.
}
\label{fig-greenfunctions}
\end{figure*}

\subsection{Monte Carlo procedure}
We perform stochastic sampling for $\boldsymbol{\tau}$ and $\boldsymbol{\mu}$ in Eq.~(\ref{partf2}).
In order to fulfill ergodicity, we need to perform two types of updates: (i) the segment addition/removal as in the ordinary Anderson model\cite{werner-PRL-2006}, and (ii) a $U_{fc}$ addition/removal update.
For update (i), we perform random choice for a new segment
in the same way as in Ref.~\onlinecite{werner-PRL-2006}: a position of the segment is chosen from the interval $[0:\beta)$ and its length $L$ from $(0:\ell_{\rm max})$.
In the present case, the update probability $R$ for the segment addition is given by
\begin{eqnarray}
R(q \rightarrow q+1) &=&  V^2 \left(\frac{u_{1}}{u_{0}} \right)^{n} {\rm e}^{-\widetilde{\varepsilon}_{f} L}  \frac{\beta \ell_{\rm max}}{q+1} \nonumber\\
&\times&  \frac{{\rm det} 
D(\boldsymbol{\tau}^{(q+1)}, \boldsymbol{\mu})}{{\rm det}\, D(\boldsymbol{\tau}^{(q)}, \boldsymbol{\mu})},
\end{eqnarray}
where $\widetilde{\varepsilon}_{f} = \varepsilon_{f} + (\alpha_c-1/2) U_{fc}$, and 
$\boldsymbol{\tau}^{(q)}$ and $\boldsymbol{\tau}^{(q+1)}$ denotes the configurations of order $q$ and $q+1$, respectively.
The factor $(u_1/u_0)^n$ accounts for the change of the time-dependent potential due to the newly inserted segment, where $n$ is the number of $\mu_{\ell}$ which are located on the inserted segment.
For update (ii), suppose that we try to add $U_{fc}$ term at time $\mu$ which is randomly chosen in the interval $[0:\beta)$. 
The update probability $R$ is given by
\begin{eqnarray}
R(m \rightarrow m+1) &=& (-1) u(\mu) \frac{\beta}{m+1}\nonumber\\
&\times&
 \frac{{\rm det}\, D(\boldsymbol{\tau}, \boldsymbol{\mu}^{(m+1)})}{{\rm det}\, D(\boldsymbol{\tau}, \boldsymbol{\mu}^{(m)})}.
\label{prob-ufc}
\end{eqnarray}
Here $\boldsymbol{\mu}^{(m)}$ and $\boldsymbol{\mu}^{(m+1)}$ denotes the configurations of order $m$ and $m+1$, respectively.

A comment on the technical parameters, $\alpha_f$ and $\alpha_c$, is now in order.
The value of $\alpha_f$ is first determined so that the potential $u(\tau)$ in Eq.~(\ref{time-dep-pot}) does not change the sign: we choose $u(\tau)>0$. At the same time, $u(\tau)$ should be small because large values results in a high expansion order.
These conditions lead to $u_{0}=\delta$, $u_{1}=U_{fc}+\delta$ for $U_{fc}>0$, and $u_{0}=-U_{fc}+\delta$, $u_{1}=\delta$ for $U_{fc}<0$ with $\delta$ being a small positive value,\cite{rubtsov-2005} e.g., $\delta=0.01$. 
Thus, $\alpha_{f}=-\delta/U_{fc}$ for $U_{fc}>0$ and $\alpha_{f}=1-\delta/U_{fc}$ for $U_{fc}<0$ in Eq.~(\ref{eq:tildehfc}).
The parameter $\alpha_c$ is next determined from a condition for positive weight. Noting $u(\tau)>0$, Eq.~(\ref{partf2}) gives $(-1)^{m}\det D>0$ (see also (\ref{prob-ufc})).
By considering a configuration with $m=1$, we obtain the condition $\tilde{g}(-0)-\alpha_c<0$, which leads to $\alpha_c>1$ since $0<\tilde{g}(-0)<1$.

Figure~\ref{fig-prob} shows probability distributions $P_{q}$ ($P_{m}$) of the expansion order for $V$ ($U_{fc}$), and its average $\langle q \rangle = \sum_q q P_q$ ($\langle m \rangle = \sum_m m P_m$).
Here, $P_q$ is defined by $P_{q} = {\cal Z}^{-1} \sum_m {\cal Z}_{q,m}$ with 
the partition function ${\cal Z}=\sum_{q,m} {\cal Z}_{q,m}$ in Eq.~(\ref{partf2}),
and $P_{m} = {\cal Z}^{-1} \sum_q {\cal Z}_{q,m}$.
The quantity $\langle q+m \rangle$, which corresponds to the average matrix size, determines the computational time. It turns out from the inset of Fig.~\ref{fig-prob} that we can reach up to $|U_{fc}| \sim 2$ in this parameter set.
The matrix size is proportional to $\beta$ and the computable range of $U_{fc}$ gets narrower as temperature decreases.

\subsection{Green's functions}
We present how to compute the single-particle Green's functions. 
In the present system with $U_{fc}$, the self-energy has the off-diagonal component $\Sigma_{fc}(i\omega_n)$ between $f$ and $c$ as well as the diagonal components, $\Sigma_{f}(i\omega_n)$ and $\Sigma_{c}(i\omega_n)$.
Hence, it is convenient to express the Green's functions in the real space.
The impurity-site Green's functions are written as
\begin{align}
\begin{pmatrix}
G_{c} & G_{cf} \\
G_{fc} & G_{f}
\end{pmatrix}^{-1}
= \begin{pmatrix}
g & 0 \\
0 & g_f
\end{pmatrix}^{-1}
- \begin{pmatrix}
\Sigma_{c} & V+\Sigma_{cf} \\
V+\Sigma_{fc} & \Sigma_{f}
\end{pmatrix},
\end{align}
where $g_f=(i\omega_n-\varepsilon_f)^{-1}$ and $\Sigma_{cf}(i\omega_n)=\Sigma_{fc}(i\omega_n)^*$.
The energy argument was omitted for simplicity.
Solving this equation, we obtain explicit expressions for the Green's functions. The $f$ component, for example, is evaluated to give
\begin{align}
G_f &= \left[ i\omega_n-\varepsilon_f - \Sigma_f - \frac{|V+\Sigma_{fc}|^2}{g_{c}^{-1} - \Sigma_c} \right]^{-1}.
\end{align}
A difference to $G_f$ in the ordinary Anderson model is the renormalization of the hybridization $V \to V+\Sigma_{fc}$ and the correction by $\Sigma_c$.
To obtain full information, we need to evaluate three quantities in the present system.

In the simulation, we compute the following quantity in the imaginary-time domain:
\begin{eqnarray}
F_{\gamma \gamma'}(\tau) &=&  -\frac{1}{V^2 \beta}  \nonumber\\
&\times&
 \left\langle 
\sum_{j (\gamma)} \sum_{i (\gamma')} 
\left[ D(\boldsymbol{\tau}, \boldsymbol{\mu})^{-1}  \right]_{ji} \Delta(\tau,\tau_{j}^{\prime}-\tau_{i})  \right\rangle_{\rm MC},\nonumber\\
\end{eqnarray}
where $\gamma = f, c$. The range of the summation depends on $\gamma$: 
$\sum_{i(f)} \equiv \sum_{i=1}^{q}$ and $\sum_{i(c)} \equiv \sum_{i=1}^{q+m}$.
The function $\Delta$ is defined in Ref.~\onlinecite{werner-PRL-2006}.
After the Fourier transform, $F_{\gamma \gamma'}(i\omega_n)$ yields the Green's functions by
\begin{align}
G_f &= F_{ff}, \\
G_c &= \tilde{g} + \tilde{g} V F_{cc} V \tilde{g}, \\
G_{fc} &= F_{fc} V \tilde{g},
\quad
G_{cf} = \tilde{g} V F_{cf}.
\end{align}
We may use the relation $G_{cf}(i\omega_n)=G_{fc}(i\omega_n)^*$ to improve the accuracy.
When $U_{fc}=0$, i.e., in the non-interacting Anderson model, $F_{\gamma \gamma'}$ is independent of the indices. In this case, the above formulas are reduced to the ordinary relations 
$G_{c}=g+gVG_{f}Vg$ and $G_{cf}=g V G_{f}$.

In order to confirm validity of our algorithm, we solve a toy model with a single conduction-electron site, i.e.,  the model~(\ref{eq-ham-0}) with ${\cal H}_{c}$ replaced by ${\cal H}_{c}^{\rm toy}=\varepsilon_{c} c^{\dag} c$. 
This model can be solved by diagonalization of a 4 $\times$ 4 matrix.
Figure~\ref{fig-greenfunctions} shows the Green's functions, $G_f$, $G_c$ and $G_{fc}$, computed in the CT-QMC, compared with the exact results. 
The error bars are smaller than the point size. We can see complete agreement between the CT-QMC and the exact results.

To obtain spectrum from the Matsubara Green's functions, we perform analytical continuation $i\omega_n \to \omega + i\delta$ by the Pad\'e approximation.
Although this approximation can not be completely controlled in general, the data obtained in the CT-QMC simulation are highly accurate 
so that this simplest method gives reasonable spectra.
To enforce the particle-hole symmetry, we dropped the real part of the Green's function $G_{f}(i\omega_{n})$ that comes from statistical errors.

\begin{figure}
\centering
\includegraphics[width=0.9\hsize]{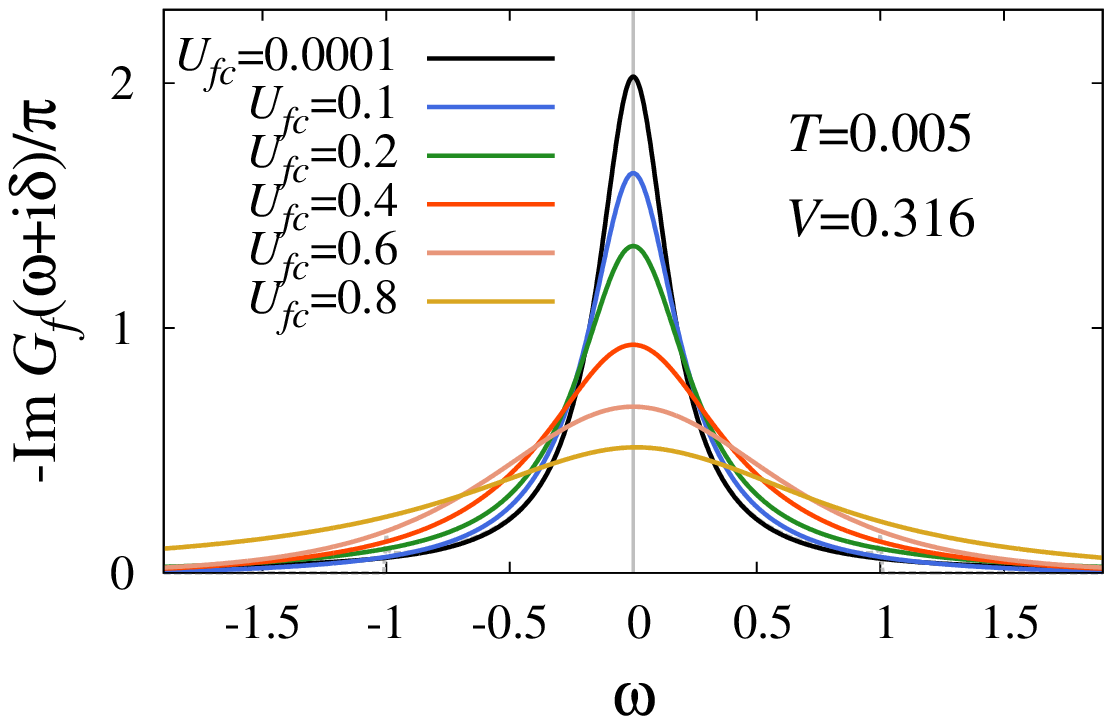}
\includegraphics[width=0.9\hsize]{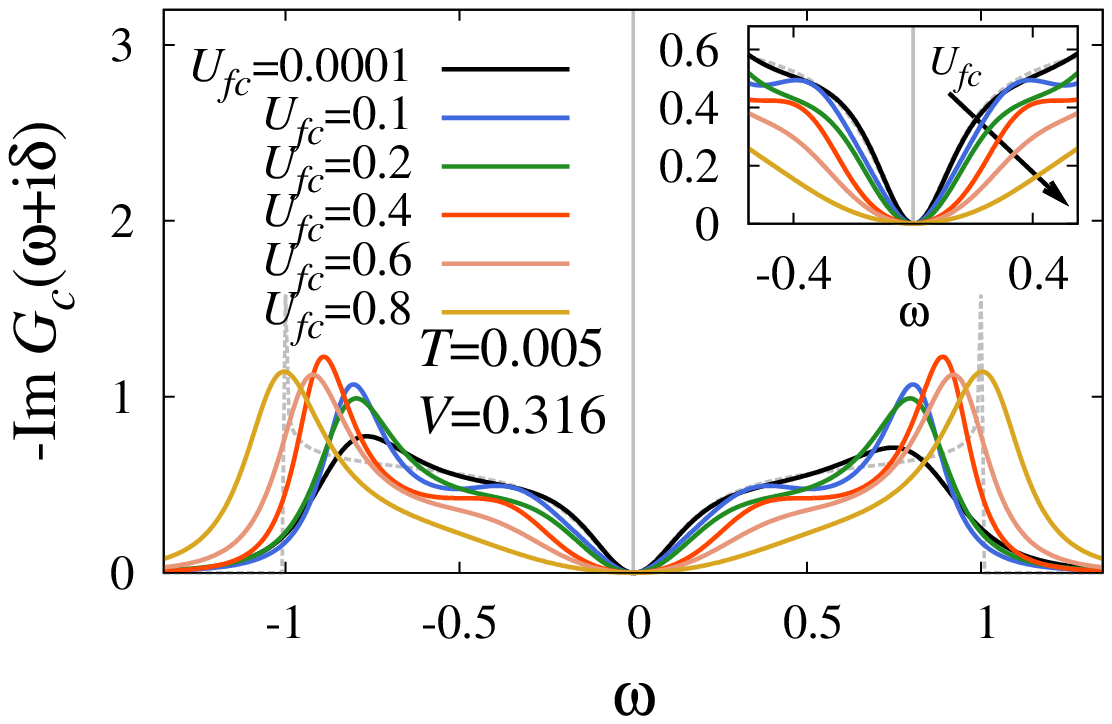}
\caption{(Color online) $f$-electron and conduction electron density of states in the real-frequency domain for positive values of $U_{fc}$.
The exact analytic result for $U_{fc}=0$ by use of Eq.~(\ref{analytic})
is also shown as {\sl dashed gray curve}. The {\sl inset} shows the enlarged region around $\omega = 0$.
The parameters are chosen as $V=0.316$ and 
$T=0.005$.
}
\label{fig-gfcdos-pUfc}
\end{figure}

\section{Numerical results}\label{section-numresults}

\subsection{Single-particle spectra of conduction and local electrons}

\begin{figure}
\centering
\includegraphics[width=0.9\hsize]{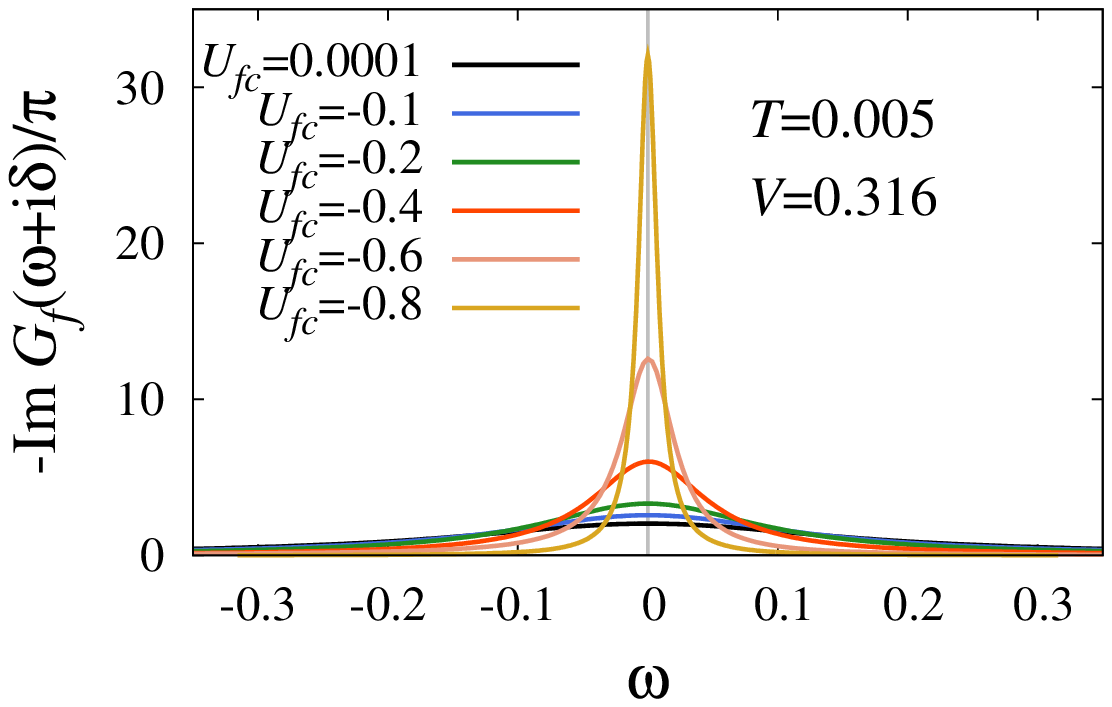}
\includegraphics[width=0.9\hsize]{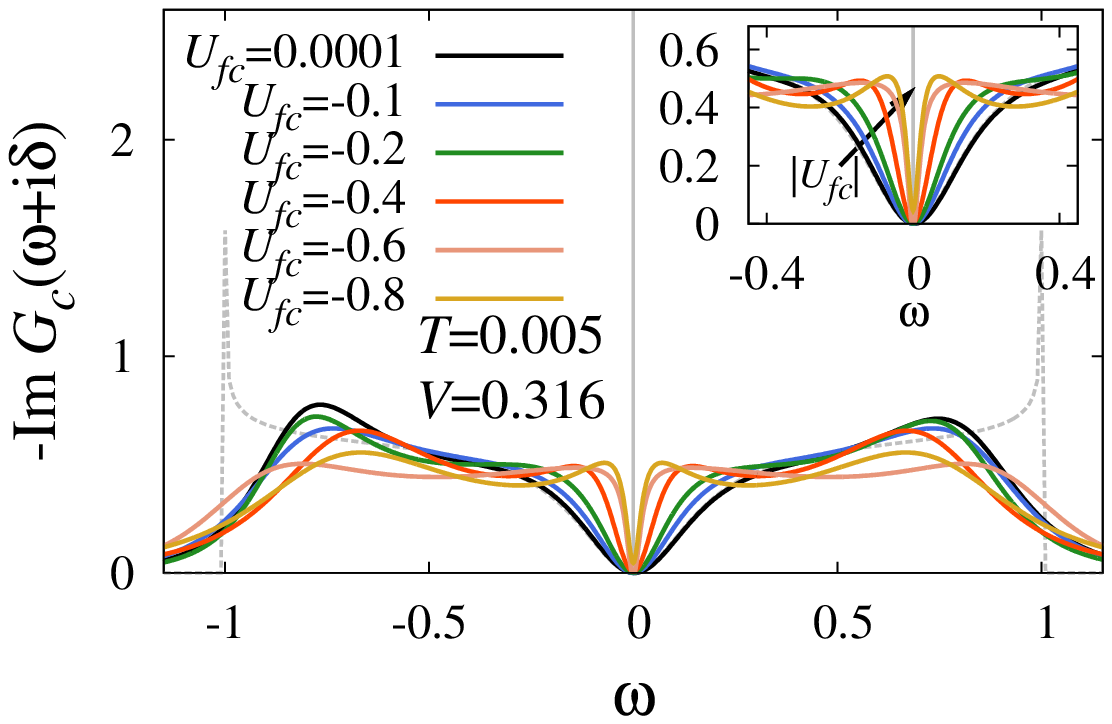}
\caption{(Color online) $f$-electron and conduction electron density of states in the real-frequency domain for negative values of $U_{fc}$.
The exact analytic result for $U_{fc}=0$ by use of Eq.~(\ref{analytic}) is also shown 
as {\sl dashed gray curve}. The {\sl inset} shows the enlarged region around $\omega = 0$.
The parameters are chosen as $V=0.316$
and $T=0.005$.}
\label{fig-gfcdos-mUfc}
\end{figure}

The single electron spectra 
$\rho_f(\omega)=-{\rm Im}\, G_{f}(\omega+i\delta)/\pi$ and
$\rho_c(\omega)=-{\rm Im}\, G_{c}(\omega+i\delta)/\pi$ 
with $\delta$ being positive infinitesimal 
are shown for $U_{fc}>0$
(Fig.~\ref{fig-gfcdos-pUfc}) and $U_{fc}<0$
(Fig.~\ref{fig-gfcdos-mUfc}) of 
the Coulomb interaction 
at finite temperature.
We use a constant density of states for the conduction electrons in the simulation as
\begin{eqnarray}
\rho_{0}(\varepsilon) = \frac{1}{2D} \Theta(D - |\varepsilon|),\label{eq:dos}
\end{eqnarray}
where we set $D=1$ as the unit of energy.
The $f$-electron resonance width increases with increasing values of $U_{fc}$ in the positive range, while decreases in the negative range, 
which is consistent with renormalized hybridization.

We find the reduction of the conduction electron density of states at the Fermi energy compared to the non-interacting density of state $\rho_{0}$, which property was already recognized long time ago\cite{mezei-1971}.
Namely, the conduction electron density of states can be expressed as
\begin{eqnarray}
\rho_{c}(z) = 
-\frac{1}{\pi}{\rm Im} \left[ g_{c}(z) + g_{c}(z)t(z) g_{c}(z)  \right]
\label{eq-rhoc}
\end{eqnarray} 
with $z=\omega+i\delta$.
We 
can use the approximation $g_{c}(0)=-i\pi \rho_{0}$ around the Fermi level and the $t$-matrix can be expressed as $t(0)=-i \sin^2 \delta / (\pi \rho_{0})$. Thus, we obtain from Eq.~(\ref{eq-rhoc}) that 
\begin{eqnarray}
\rho_{c}(0) = \rho_{0} (1 - \sin^{2} \delta) = \rho_{0} \cos^2 \delta.\label{eq-roc2}
\end{eqnarray} 
Since the phase shift $\delta_{U}$ coming from the $U_{fc}$ process is zero at the Fermi energy (see Eq.~(\ref{eq-phsufc})), we have $\delta=\pi/2$ for resonant scattering, which gives the vanishing of the conduction electron density of states at the Fermi level.

In the case of $U_{fc}=0$, we obtain the analytic result:
\begin{align}
t(z) = V^2G_f(z) = V^2\left[ 
z-\frac{V^2}{2D}\ln\left( \frac{z+D}{z-D} \right)
 \right]^{-1},
\label{analytic} 
\end{align}
which gives singularity of $\rho_c(\omega)$ at $\omega = \pm D$.

\subsection{Renormalized hybridization}

\begin{figure}
\centering
\includegraphics[width=0.9\hsize]{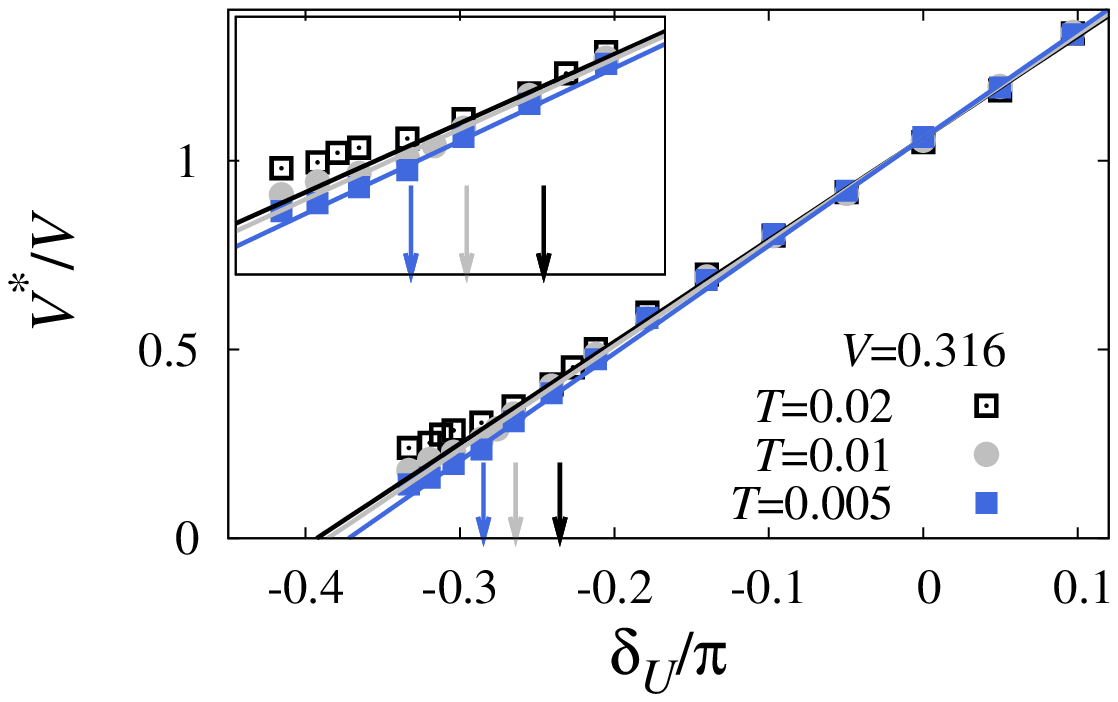}
\includegraphics[width=0.9\hsize]{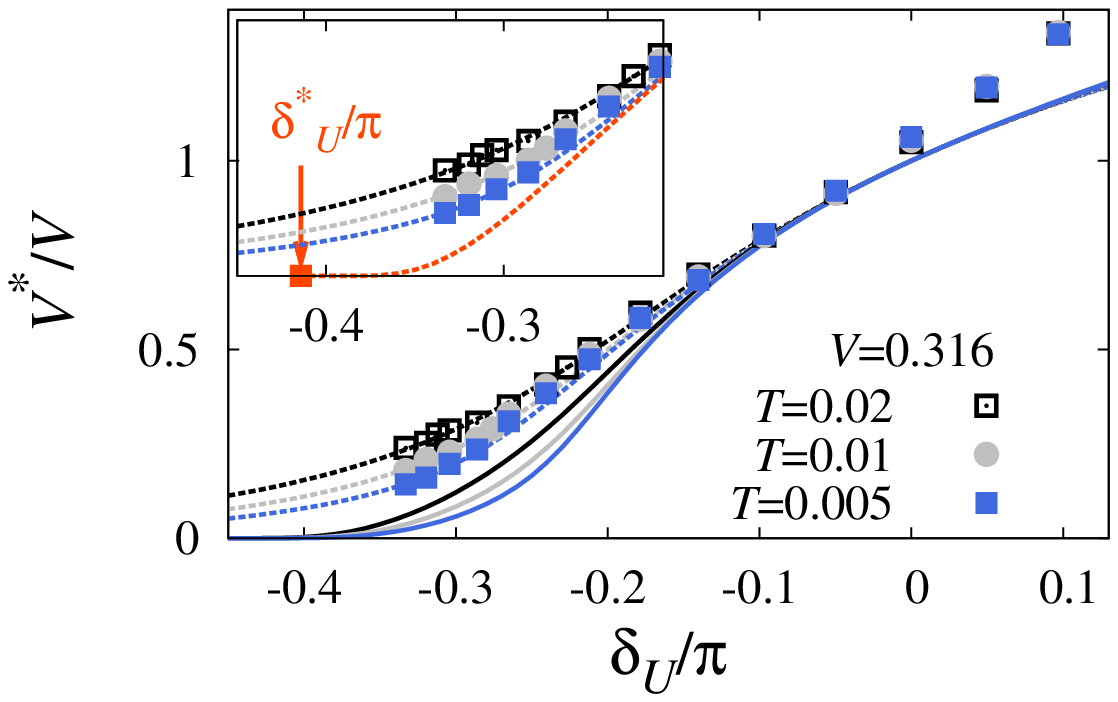}
\caption{(Color online) Renormalized hybridization obtained in the simulation at different temperatures.
The top panel shows a linear fitting 
against the phase shift $\delta_{U}$, 
where the arrows show the characteristic value $\delta_{U}^{\ast}$ at different temperatures.
The bottom panel compares the numerical data with the perturbative RG ({\sl solid line}) and the phase shift scheme ({\sl dashed line})
as a function of $\delta_{U}/\pi$.
The {\sl inset} enlarges  the renormalized hybridization 
near 
a quantum critical point at $T=0$.
The red line shows the result obtained from Eq.~(\ref{V^*}) together with
$\delta_{U}^{\ast}/\pi$.
The bare hybridization is chosen as $V=0.316$.
}
\label{fig-gfcdos-renV1}
\end{figure}

The $U_{fc}$ dependence of the renormalized hybridization can be quantitatively obtained from the single particle spectra shown in Figs.~\ref{fig-gfcdos-pUfc} and \ref{fig-gfcdos-mUfc}.
Namely, we fit the spectrum by the Lorentzian, and deduce the width $\Delta^*(T)=\pi \rho_0 (V^{\ast})^2$ and the renormalized hybridization $V^*$.
The result is summarized in Fig.~\ref{fig-gfcdos-renV1} at different temperatures as a function of the phase shift  $\delta_{U}$.
We find a linear dependence of $V^{\ast}$ on $\delta_{U}$ around the 
noninteracting limit of $\delta_{U}=0$ (namely $U_{fc}=0$) as it is shown in the left part of Fig.~\ref{fig-gfcdos-renV1}. 
For small absolute values of $\tilde{u}=\delta_U/\pi$, the linear dependence should follow from Eq.~(\ref{V^*}):
\begin{align}
\frac{V^*}V \sim \left( \frac\Delta D \right)^{-\tilde{u}} \sim 
1 +\tilde{u}\ln\left( \frac D\Delta \right),
\end{align}
which explains semiquantitatively the behavior around $\delta_U/\pi \sim 0$.
Actually, the linear dependence prevails in a wide range of $\delta_{U} \ (<0)$ down to about $\delta_{U}^{\ast}$ given by Eq.~(\ref{delta_U})

Next, we try to fit the numerical data with the perturbative RG result by taking $u=\rho_{0}U_{fc}$ in the finite temperature expression of the renormalized hybridization given in Eq.~(\ref{eq-deltapt}), and also with the phase shift scheme by taking $u\, \rightarrow \, \tilde{u}={\rm arctan} (\pi \rho_{0} U_{fc})/\pi$.
The fits are shown in the bottom part of Fig.~\ref{fig-gfcdos-renV1}.
We find that the phase shift scheme works well in the range of $|\delta_{U}|>|\delta_{U}^{\ast}|$, i.e. where the linear fit breaks down. 
The perturbative RG description does not work except for a narrow range of $U_{fc}$ in the vicinity of $\delta_{U}=0$ (namely for $U_{fc}=0$).

We note that numerical results for the renormalized hybridization $V^{\star}/V$ given in Fig.~\ref{fig-gfcdos-renV1} shows slight deviation from unity at $U_{fc}=0$.
This is due to numerical inaccuracy of the simulation with 
the large value of $V=0.316$, which is  
comparable to the half-bandwidth $D=1$.
We have checked that this deviation decreases
by decreasing $V\ (>0)$ in the calculation.

\begin{figure}
\centering
\includegraphics[width=0.9\hsize]{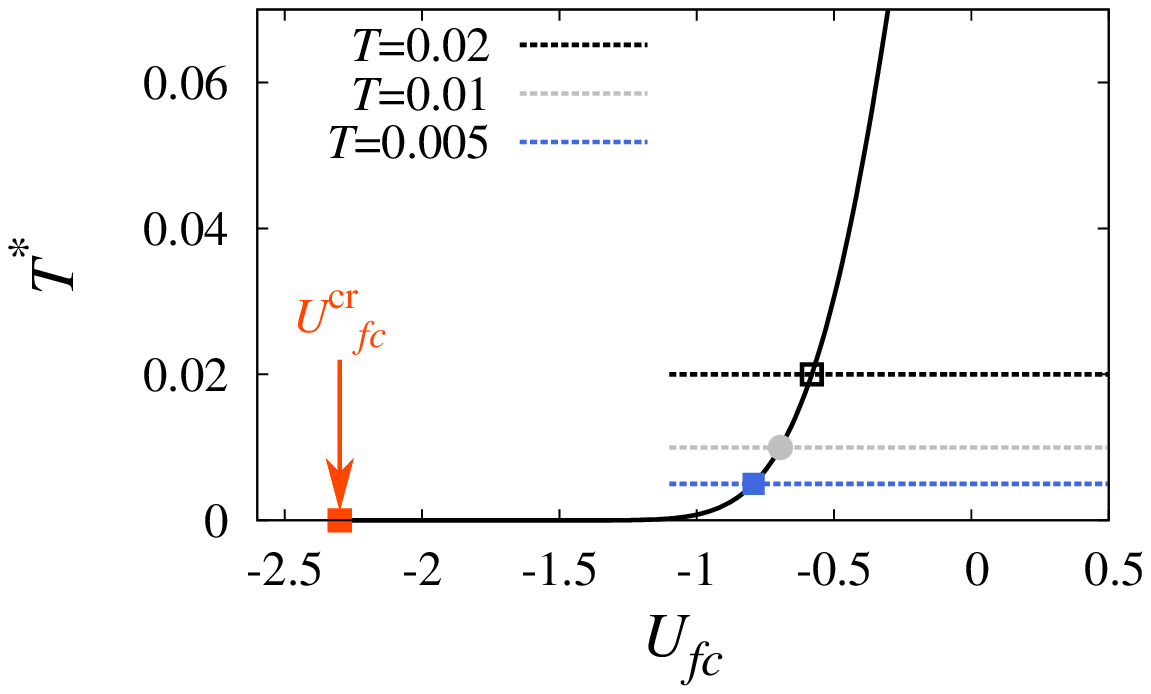}
\includegraphics[width=0.9\hsize]{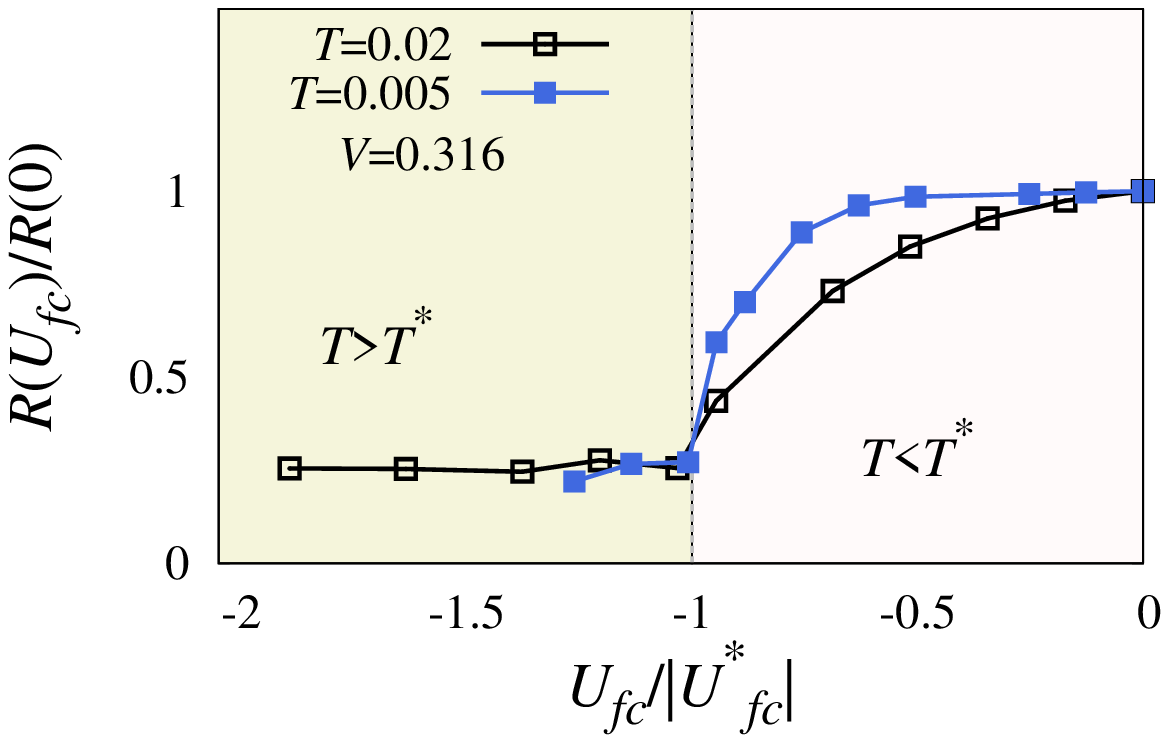}
\caption{(Color online) {\sl Top panel:} Characteristic temperature $T^{\ast}$ as a function of $U_{fc}$. {\sl Dashed lines} show the temperature values used in the numerical calculations. {\sl Bottom panel:} $U_{fc}$ dependence of the electric resistivity at temperatures $T=0.02$ and $T=0.005$.
The bare hybridization is chosen as $V=0.316$.
}
\label{fig-resistivity}
\end{figure}

In the left part of Fig.~\ref{fig-resistivity} the characteristic temperature $T^{\ast}=\Delta^*$ 
is shown as a function of $U_{fc}$ together with the temperature values we used in the simulation ({\sl dashed lines}).
The intersection of these horizontal lines with the curve of $T^{\ast}$ gives the characteristic Coulomb interaction at the different temperatures as $U_{fc}^{\ast}(T=0.02)=-0.581$, $U_{fc}^{\ast}(T=0.01)=-0.695$, and $U_{fc}^{\ast}(T=0.005)=-0.793$ (see {\sl dot symbols}) by taking $V=0.316$.
The temperature dependence of the density of states persists 
more and more to lower $T$ as we come closer to the quantum critical point.  In other words, the characteristic temperature becomes tiny.  
Hence it is difficult to identify $U_{fc}^{\rm cr}$ precisely in our simulation.  In the range
$U_{fc} < U_{fc}^{\rm cr}$, we no longer observe a smooth peak in $\rho_f(\omega)$ around $\omega = 0$.  The simulation does not converge well in this range.

To demonstrate the drastic change around $T=T^*$ and $U_{fc} =U_{fc}^{\ast}(T)$, we calculate
the resistivity $R(T)$ using the formula
\begin{eqnarray}
R(T)^{-1} = 
\int_{-\infty}^{\infty} d \varepsilon \left( - \frac{\partial f (\varepsilon)}{\partial \varepsilon} \right) \tau(\varepsilon),  
\end{eqnarray}
where $f(\varepsilon)$ is the Fermi function and the relaxation time $\tau(\varepsilon)$ is obtained from the real-frequency $t$-matrix as
\begin{eqnarray}
\tau(\varepsilon)^{-1} =  -2 {\rm Im} \, t(\varepsilon).
\end{eqnarray}
Figure~\ref{fig-resistivity} shows the resistivity obtained  as a function of the Coulomb interaction at different temperatures.
We find a distinct change in the resistivity at a given temperature $T$ almost exactly at  $U_{fc}=U_{fc}^{\ast}(T)$, where $U_{fc}^{\ast}$ is the characteristic Coulomb interaction given in Eq.~(\ref{eq-charufc}).
The resistivity shows substantial $U_{fc}$ dependence in the range of $|U_{fc}|<|U_{fc}^{\ast}|$, i.e. where $T<T^{\ast}$, 
while the resistivity is almost independent of $U_{fc}$ for $|U_{fc}|>|U_{fc}^{\ast}|$, i.e. where $T>T^{\ast}$. 
Thus, we confirm that $U_{fc}^{\ast}(T)$ as calculated in the phase shift scheme separates the two regimes with different behaviors of the resistivity.

\section{
Summary and
Discussion }\label{section-discussion}

In this paper we have studied the single-channel interacting resonant level model in a wide range of $U_{fc}$ under finite hybridization $V$ by exact numerical method.
As theory, we rederived the renormalized hybridization by perturbative RG approach which is valid for small values of the bare parameters $V$ and $U_{fc}$, and took also the phase shift scheme that includes $V$ in the lowest order but $U_{fc}$ up to infinite order.
The derived dynamics and thermodynamics are compared with the results of perturbative RG and phase shift scheme to check their applicability.
We find an excellent agreement with the phase shift scheme in the negative range of $U_{fc}$ including a quantum critical point at $U_{fc}^{\rm cr} =  -2\tan(\sqrt{2}\pi)/\pi \approx -2.3$ since the renormalized hybridization becomes negligible in this range. As the Coulomb interaction is increased in the positive range, however, the numerical results highly deviate from the scaling result.
By calculating physical quantities such as electric resistivity at finite temperatures, we demonstrate the change around the characteristic energy realized as crossover to the ground state.

Now we discuss possible relation of the present results to 
the unusual heavy fermion state in actual systems.
We have paid special attention to the negative range of $U_{fc}$ since 
the quantum critical point with vanishing hybridization emerges in the negative range. 
We point out a possibility that the effective Coulomb interaction may be renormalized to negative value 
by interaction with phonons, for example.
Then, a possible scenario to explain the peculiar heavy fermion state of SmOs$_{4}$Sb$_{12}$ is that the 
system is close to the quantum critical point where the reduced effective hybridization gives rise to huge effective mass.

Another possible scenario is the presence of multi-channels for the conduction bands.  Then a non-Fermi liquid fixed point can emerge even for $U_{fc}>0$.
The origin of multi-channels can be either purely electronic\cite{giamarchi-1993} or because of the assistance of phonons\cite{ueda-2010}.
In real systems, however,  the condition $\varepsilon_{f} = 0$ we assumed in this paper is not satisfied in general. 
Then, like the Zeeman field in the spin Kondo case, a finite value of $\varepsilon_{f}$ acts as an external field against the charge Kondo effect. 
In this way the non-Fermi liquid fixed point becomes
unstable with a finite value of $\varepsilon_{f}$ in actual multi-channel systems,
and therefore the ground state remains a Fermi liquid with strongly enhanced effective mass. 
We will investigate the multi-channel IRLM using the accurate CT-QMC in a subsequent paper.

\acknowledgements

We are grateful to Dr. S. Hoshino for enlightening discussions.
This work is supported by the Marie Curie
Grants PIRG-GA-2010-276834 and the Hungarian Scientific Research Funds No. K106047.

\end{document}